\documentclass{article}

\usepackage{arxiv}

\usepackage[utf8]{inputenc} 
\usepackage[T1]{fontenc}    
\usepackage{hyperref}       
\usepackage{url}            
\usepackage{booktabs}       
\usepackage{amsfonts}       
\usepackage{nicefrac}       
\usepackage{microtype}      
\usepackage{lipsum}		

\usepackage{bigstrut}
\usepackage{subfigure}
\usepackage{setspace}
\usepackage{graphicx}

\usepackage{booktabs}

\usepackage{natbib} 
\usepackage{lscape}

\newcommand{\captionfonts}{}
\usepackage{xr}

\makeatletter  
\long\def\@makecaption#1#2{%
  \vskip\abovecaptionskip
  \sbox\@tempboxa{{\captionfonts #1:  #2}}%
  \ifdim \wd\@tempboxa >\hsize
    {\captionfonts   #1. #2  \par}
  \else
  {\captionfonts  \begin{center} \textbf{#1: } \textbf{#2} \end{center} \par}
  \fi
  \vskip\belowcaptionskip}
\makeatother

\title{Artificial Intelligence Alter Egos:\\
	Who benefits from Robo-investing?
	\author{ Catherine D'Hondt,
		Rudy De Winne,
		Eric Ghysels, 
		Steve Raymond
	}
}
\begin{document}
\maketitle

\thispagestyle{empty}


\onehalfspacing
\begin{center}
	--------------------------------------------------------------------------------------------------------------
\end{center}
\begin{abstract}
	\noindent
Artificial intelligence, or AI, enhancements are increasingly shaping our daily lives. Financial decision-making is no exception to this. We introduce the notion of AI Alter Egos, which are shadow robo-investors, and use a unique data set covering brokerage accounts for a large cross-section of investors over a sample from January 2003 to March 2012, which includes the 2008 financial crisis, to assess the benefits of robo-investing. We have detailed investor characteristics and records of all trades.  Our data set consists of investors typically targeted for robo-advising. We explore robo-investing strategies commonly used in the industry, including some involving advanced machine learning methods. The man versus machine comparison allows us to shed light on potential benefits the emerging robo-advising industry may provide to certain segments of the population, such as low income and/or high risk averse investors.
	
	\vskip10pt
\end{abstract}

\begin{center}
	--------------------------------------------------------------------------------------------------------------
\end{center}

\clearpage \pagestyle{plain} \addtocounter{page}{-1}
\setstretch{1.22}  \doublespacing

\newpage

\section{Introduction}

To assess the benefits of robo-investing we use a unique data set covering brokerage accounts for a large cross-section of 22,972 individual investors covering a sample from January 2003 to March 2012, and therefore includes the 2008 financial crisis. We have records of all trades, and in addition have detailed information about each individual investor's characteristics such as age, gender, education, annual net income, and most importantly, risk aversion assessed on the basis of responses to survey questions.  Although we work with Belgian individual investors, most of their trading activities pertain to foreign stocks (86\% are non-Belgian and roughly a quarter are US). Hence, our analysis pertains to international portfolio selection of stocks and ETFs.

To the best of our knowledge there has not been any assessment of the potential benefits of robo-investing over a long period of time for a heterogeneous panel of individual investors. We explore robo-investing strategies commonly used in the industry, including some involving advanced machine learning methods. The man versus machine comparison allows us to shed light on potential benefits the emerging robo-advizing industry may provide to certain targeted segments of the population, such as low income and/or investors with relatively little financial literacy.\footnote{In the US, robo-advisor start-ups saw an eight-fold increase in their AUM in recent years on the back of some retirement savings shifting to robo-advisor accounts. Cost advantages have been creating significant momentum for the industry. In addition, the success of passive investment strategies in recent years has also been beneficial.  It is therefore fair to say that robo-advisors are posing a challenge to traditional financial advisory services. One expects that some robo-advisory start-ups will probably end up in partnerships or be the subject of takeovers by established asset management firms or banks in the coming years. Moreover, the traditional asset managers themselves are also adopting robo-investing strategies. In that respect, robo-advising will become more mainstream.}

Our sample has a number of appealing features to study robo-investing. Many investment brokerage firms are now targeting individuals with modest savings as it is generally believed that smaller investors don't get the investment advice they need. In fact, 71\% or almost 90 million American families have investment account balances worth less than \$100,000. The growth of automated investment advisory services is filling a need for such investors. Our data set consists of individual investors typically targeted by robo-advising. In terms of annual net income, approximately 70\% of the investors in our sample declare an income between 20,000 and 75,000 euros. The mean portfolio value in our sample is 29,244 euros and the average investor is about 48 years old. 

Note that our paper does not directly address the effect on wealth management of adopting robo-advising, as studied by for example \cite{d2019promises}. On the one hand, our data is richer in terms of details regarding the characteristics - such as income, education, gender, risk aversion, trading habits - for each individual investor. On the other hand, we study a sample where robo-advising was not adopted by the brokerage firm whose trading data we examine. Instead, we introduce the idea of shadow robo-investors to assess the potential benefits of robo-advising. Namely, we study various robo-investors that shadow the individuals in our data set and the novelty of our approach is that we know what the investors have done in reality versus what a robo-investor would have done instead. In that sense our analysis is a real-time experiment with real data. 

Robo-investors are limited to the set of stocks and ETFs in each individual investor's history of trading - using a rolling 2-year sample.\footnote{The majority of trading occurs in either equity or ETFs as described in detail in the Appendix, see \cite{DDGR_SSRN}.} This constraint ties each robot to a specific investor in our sample via their trading history. Note that the robo-investors use {\it all} the stocks/ETFs individual investor $i$ held in the past two years, but may have sold in the meantime. Hence, the rationale is that the investor knows about the stocks/ETFs held by the shadow robo-investor. We call these shadow robo-investors Artificial Intelligence Alter Egos, or AI Alter Egos.\footnote{Since the robo-investor schemes go beyond machine learning, as they involve portfolio allocation rules, we use the more general term of artificial intelligence. In our case the AI pertains a set of computer-driven self-learning rules which determine portfolio allocations.}  

The notion of AI Alter Egos is not unique to finance, although we might be the first to coin the term.  To illustrate, let’s look at machine learning (ML) advances in other fields, such as literature and music. Today, a ML text mining algorithm can analyze the writings of a famous author and create entirely new literature in the style of the writer it was exposed to and trained on. The same can be done with music. For example, Franz Schubert started his Unfinished Symphony in B minor in 1882, but wrote only two complete movements, though he lived  another six years. Now, deep-learning ML has produced a completed version of the entire symphony. We can characterize this as Schubert's AI Alter Ego composing a new score. Would Schubert have done better than his AI Alter Ego? We prefer to leave that debate to the musicologists, but it’s fair to say it would probably be hard to address the question. Fortunately, it’s much easier to apply the notion of AI Alter Egos in a setting where comparing the outcomes of human and AI alternatives is more straightforward – such as in financial investments.

We consider three investment strategies. Two are based on a \cite{markowitz1952portfolio} mean-variance (MV) scheme and a third is based on \cite{demiguel2007optimal} involving the $1/N_{it}$ scheme where $N_{it}$ is the number of stocks held by investor $i$ over a 2-year trailing sample up to time $t.$ 
The two MV strategies differ in terms of the sophistication regarding the conditional mean and variance estimates. The first involves two-year rolling sample estimates for both the mean and variance. For the second we rev up the robot engines and replace the rolling sample estimators by respectively expected return predictions using machine learning algorithms and sophisticated conditional covariance estimators. More specifically for the conditional mean we use Elastic-Net, Random Forest, Neural Network, and model ensemble estimators. For the conditional covariance matrix - looking at a total of 683 stocks and 393 ETFs - we use the \cite{engle2017large} nonlinear shrinkage method derived from random matrix theory to correct in-sample biases of sample eigenvalues. Finally, it is important to note that robo-investors have the option to hold cash, i.e.\ decide to avoid market risk exposure. No short selling is allowed, however.

We study three rebalancing schemes: once a year, quarterly and monthly. In the main body of the paper we focus exclusively on the quarterly rebalancing scheme. Note that robo-investors buy and hold at fixed sampling frequencies - end of quarter in the lead example. This is in contrast to the individual investors in our sample who execute their trades at any point in time. 

Overall our findings are as follows. The AI Alter Ego robo-investors involving equal weighting or rolling sample mean and variance estimates perform poorly and are of little value to any of our investors.\footnote{Note that among the existing robo-investor practices there are number which proclaim using MV allocations and most likely use some type of rolling sample scheme - although most white papers are rather vague on the actual implementation.} In contrast the machine learning MV AI Alter Egos result in significant investment portfolio performance improvements for certain types of investors. In particular, those featuring high risk aversion benefit greatly from following the robo-investor strategies. Low income (low education) investors  typically also gain from the AI advise. These results confirm the claims made by practitioners in the industry regarding the promises the use of AI hold for the future of the FinTech industry. More intriguing, and somewhat unexpected are our results pertaining to the performance during the financial crisis. Robo-investors outperform a large swath of investors. In fact, the median robo-investor moves into cash (because of negative expected returns using AI) whereas individuals feature behavioral biases, such as the disposition effect (cfr.\ \cite{odean1998investors}) with unfortunate consequences during the onset of the financial crisis.

As a by-product of our analysis, we also identify which machine learning methods perform well. While deep learning is often the best across a large cross-section of stocks, a close second-best is a much simpler linear prediction model with elastic net penalty based on the same set of predictor, namely those suggested by  \cite{welch2007comprehensive}, which consist of a mixture of firm-specific and macroeconomic covariates. Put differently, the gains from non-linear models is marginal at best.

The paper is organized as follows. In section \ref{sec:brokerage} we describe the brokerage data, with some of the details appearing in the Appendix.\footnote{Further details see \cite{DDGR_SSRN}.} Section \ref{sec:robo-inv} describes the various robo-investor schemes. Section \ref{sec:empres} reports the empirical results. Section \ref{sec:concl} concludes the paper.

\section{A Large Panel of Individual Brokerage Accounts \label{sec:brokerage}}

Our primary data set comes from a large Belgian online brokerage firm and consists of the trading accounts of 22,972 individual investors. This unique data spans about 10 years from January 2003 to March 2012, and therefore includes the 2008 financial crisis. We have detailed information about each trade, such as the instrument, the time-stamp, the trade direction, the executed quantity, the trade price, and explicit transaction costs. The details of the data are described in Appendix pf \cite{DDGR_SSRN}. We focus on common stock investments as well as ETFs and exclude other financial instruments.\footnote{In Appendix we document that 6,741 investors also traded options and warrants with an aggregate number of 602,833 trades and 6,665 investors traded mutual funds with an aggregate number of 260,120 trades. Only a few investors (i.e.\ 1,813) traded bonds with an aggregate number of 5,999 trades.} Trading of ETFs, mutual funds,  options and warrants is more prevalent with high income/education investors. Trading of bonds is overall insignificant. Because we examine robo-advisors which are mean-variance investors we focus exclusively on stocks and ETFs which best fit the portfolio allocation model. For high income/education investors in particular this means we leave out to a certain degree other assets which we have available.  After applying some filters described in the Appendix, we end up with a sample of 1,590,199 (stocks) + 60,344 (ETFs) = 1,650,543 trades (and more than 13 billion euros traded in stocks and close to 1 billion euros in ETFs) over the 111-month period covering 683 stocks and 393 ETFs or 70\% of all the investors' trading activity.

Using the trading data we build end-of-month portfolios for each investor and use historical market data to compute monthly portfolio market values. We also compute both monthly and daily returns. Combining end-of-month portfolio market values with the corresponding monthly aggregate cash-flows, we calculate for each investor 110 (i.e.\ from February 2003 to March 2012) monthly portfolio gross and net returns (the latter net of transaction costs). Investors are included as robo-investor candidates if they satisfy the return criteria - sufficient time periods and returns with no extraordinary outliers -  and minimum trade restrictions (see \cite{DDGR_SSRN} Appendix for further details). In particular, we drop investors with more than 106 missing values in their return series (i.e.\ at least 4 months of returns are needed to keep an investor) and drop outliers as well. This decreases the sample to 20,622 investors (down from 22,972). To simplify the analysis we do not take into account transaction costs (which are available for each trade) neither for the individual investment accounts nor for the robo-investor ones. Since robo-investors trade less than the average/median investor, namely only once a quarter in the lead example,  this should yield conservative estimates of the robo-investing gains.

Our data set also includes an extensive set of individual investor characteristics, such as age, gender, education,  annual net income and a risk aversion measure  based on surveys. Although we work with Belgian individual investors,  most of their trading activities involve foreign stocks (mainly the US and bordering countries France and Germany). The majority of stocks pertain to the technology sector (16.93\%), financials (15.91\%), and industrials (14.01\%). 

As noted in the Introduction, our data set consists of individual investors typically targeted by robo-advising. We have about 70\% of the investors in our sample who declare an annual net income between 20,000 and 75,000 euros. Only a minority (3.36\%) earns more than 150,000 euros per year.\footnote{The income measure reported in our data is
	recorded once, when the investor completed the MiFID tests. The classification may therefore be noisy over the 10 year sample period, particularly for the early entries.}
The average investor is about 48 years old and executes monthly 2.76 trades across 2.05 different stocks for a volume of 18,237 euros. Consistent with the literature, investors in our sample are under-diversified; the average (median) investor holds a five-stock (three-stock) portfolio. The average end-of-month portfolio value is about 28,003 euros (with a median value of about 7,552 euros). As for risk aversion, the majority of investors seem to be risk tolerant since 65.33\% of them declare a medium risk aversion and 27.88\% of them even a low risk aversion.

In terms of performance, our investors earn an average monthly gross return of 0.42\% on stocks and ETFs (median return of 0.13\%), with a volatility of 10.04\%. This high average volatility of individual portfolio gross returns is not surprising given our sample period includes turbulent market conditions.\footnote{As detailed in \cite{DDGR_SSRN} Appendix, to calculate portfolio returns, we opt for an approximation of the Modified Dietz Method, aiming at delivering a return close to the money-weighted rate of return (e.g., \cite{Shestopaloff2007}).}

\section{Robo-Investors \label{sec:robo-inv}}

The robo-investors are limited to the set of stocks and ETFs in each individual investor's history of trading - using a rolling 2-year sample. This constraint ties each robot to a specific investor in our sample via their trading history. We call these shadow robo-investors Artificial Intelligence Alter Egos. Robo-investors have the option to hold cash, i.e.\ decide to avoid market risk exposure, but no short-selling occurs in our sample nor is it allowed for in the design of the robots. The two-year window is arguably somewhat arbitrary. Our results hold for longer windows. Shorter windows are less appealing given the trading frequency of many investors, with only a median of 2 trades per month.
The portfolio allocations of robo-investors occur at fixed intervals, either monthly, quarterly or annually. In the main body of the paper we focus exclusively on the quarterly results.\footnote{In the Online Appendix of \cite{DDGR_SSRN}, the monthly and detailed quarterly results are reported. The annual results are available on request. In the computations of returns we ignore transaction costs. Since our focus is quarterly trading frequencies this is a reasonable abstraction. The monthly robo-investor results are arguably more suspect of being overstated because transaction costs are not accounted for.} 

\subsection{AI Alter Egos}

We construct three types of AI Alter Ego robo-investors. As we already noted, each setting only uses stocks and ETFs held by an individual investor over the past two years, not the entire universe of stocks.  The table below provides two illustrative examples. When we refer to $t,$ we mean end of the year, or quarter or month, depending on the case being considered.

\bigskip
\begin{center}
{		\small
	\begin{tabular}{lllll}
		Initial & Trading t - 1 & Trading t & Investor & Robo-investor  \\
		holdings &  &  & holdings & potential holdings  \\
		& & & & \\
		Stocks 1 \& 2 & Sells all of 2 & Buys stock 3 & Stocks 1 \& 3 & Stocks 1, 2 \& 3 \\
		& & & & \\
		Stock 1 & Sells all of 1 & Buys ETF 4 & ETF 4 & Stock 1 \& ETF 4 \\  
	\end{tabular}%
}	
\end{center}

\medskip

\noindent The first line portrays an investor holding two stocks - say 1 and 2 - at time $t$ - 2 (column called Initial holdings). At the end of $t$ - 1, the investors sells all holdings of stock 2 and at the end of the subsequent period $t$ buys stock 3. Hence, at the end of $t$ she/he holds stocks 1 and 3. The robo-investor has stocks 1, 2 and 3 to form a portfolio. The second case is similar, but the investor only holds stock 1, sells all of it in $t$ - 1 and buys ETF 4 in $t.$ The robo-investors has two assets to select from. It is important to stress that the robo-investor may hold cash, i.e.\ decide not to put all the money in the stock market. This will be important as will become clear when discussing the empirical results.

To proceed, we need to introduce some notation. Let $S_{it}$ be the set of stocks/ETFs investor $i$ held over a two-year period up to time $t.$ The above illustrative examples clarified that this does not mean that the investor holds these stocks/ETFs at the end of year/quarter/month $t.$ It only means that the investor held these stocks/ETFs in the recent two-year history. We denote by $T_i$ the duration of time (months/quarters/years whichever applies) investor $i$ appears in the sample. Moreover, we denote by $N_{it}$ = $\# S_{it},$ the number of stocks/ETFs in the set. We only consider investors with $N_{it}$ $\geq$ 2 $\forall$ $t$ =  1, $\ldots,$ $T_i.$ This ensures that the investment opportunity set contains a minimally sufficient set of stocks/ETFs for the robo-investors. This leaves us with 20,622 investors who satisfy this criteria and are included in our analysis.

The robo-investors buy at the end of $t$ and hold until end of $t + 1,$ i.e.\ for a month, quarter or full year.\footnote{In between rebalancing periods, the portfolio weights adjust according to the performance of an individual asset relative to the performance of the portfolio as a whole. In particular when $t+1$ is not a rebalancing period, $w_{i,t+1}$ = $w_{i,t} (1+r_{i,t+1})/[\sum_{i=1}^N w_{i,t} (1+r_{i,t+1})]$}  We then compute holding period returns for the robo-investor, $r_{i,t+1}^{ae}$, and compute the alter-ego-less-investor's realized return spread as $r_{i,t}^{s} = r_{i,t}^{ae} - r_{i,t}.$

The first type shadows each individual investor in our sample using the \cite{demiguel2007optimal} equal weighting rule, the second and third rely on  a mean-variance \cite{markowitz1952portfolio} strategy with a short-sale constraint. The difference between the second and third variations is the sophistication of expected return and risk estimators. In the second approach a simple rolling sample estimator is involved for expected returns and the linear shrinkage estimator of \cite{ledoit2004honey} for second conditional moments. In the third case, machine learning and conditional covariance estimators are used.  More specifically for the conditional mean we use Elastic-Net, Random Forest, Neural Network, and model ensemble estimators. For the conditional covariance matrix we use the \cite{engle2017large} nonlinear shrinkage method derived from random matrix theory to correct in-sample biases of sample eigenvalues.

\paragraph{Equal Weights} We endow the robo-investor with a \cite{demiguel2007optimal} $1/N_{i,t}$ strategy. In particular, for each individual $i,$ the Alter Ego buys and holds at time $t$ all the stocks in the set $S_{it}$ with equal allocations $1/N_{it}.$ Henceforth we will refer to this as the EW portfolio rule.

\paragraph{Rolling Sample Markowitz}	
The mean-variance optimal portfolio is constructed as the maximum Sharpe ratio subject to the short-sale constraint and the individual's investment opportunity set. Investor $i$ selects from the set $S_{it}$ of stocks. Critical to the optimal portfolios are estimates of conditional expected returns ($\mu_t^i$) and the conditional covariance matrix of returns ($\Sigma_t^i$) for the stocks in the set $S_{it}.$
The robo-investor solves for $\hat w_{i,t}$ selecting among these stocks according to:
\begin{eqnarray*}
&\max_{w_{i,t}} w_{i,t}'\mu_t^i - \frac{\gamma}{2} w_{i,t}' \Sigma_t^i w_{i,t} \\
& w_{i,t} \geq 0,
\end{eqnarray*}
where $\gamma$ is often interpreted as a risk aversion parameter which we set equal to one as it maximizes the Sharpe ratio.
We estimate $\mu_t^i$ with two-year rolling-window historical averages, $\hat \mu_t^i$ = $\frac{1}{k} \sum_{j=0}^{k-1} r_{t-j}^d$, where $r_t^d$ is an $N_{it} \times 1$ vector of daily returns and $k$ is the number of days in the two-year historical sample. For covariance, we also use rolling sample estimator with linear shrinkage as in \cite{ledoit2004honey} based on daily returns over the same time span.  These estimates form our naive benchmark.

\paragraph{Machine learning and Shrinkage}	
Continuing with the Markowitz allocation scheme, we explore whether increasing the complexity of the rolling sample estimators translates into improved robo-investor performance. We assume that each investor's Alter Ego robo-investor has access to a common set of models that replace the rolling sample schemes. For expected return predictions we use machine learning algorithms applied to each of the 1076 assets (683 stocks and 393 ETFs) and the Alter Ego robo-investor picks the prediction pertaining to the stocks in the sets $S_{it}.$ More specifically for the conditional mean estimates we use Elastic-Net (\cite{zou2005regularization}), Random Forest (\cite{breiman2001random}), Neural Network \cite[Chap.\ 11]{friedman2001elements}, and model ensemble estimators  \cite[Chap.\ 16]{friedman2001elements}. For the conditional covariance matrix - looking at a total of 1076 assets - we use the \cite{engle2017large} nonlinear shrinkage method derived from random matrix theory to correct in-sample biases of sample eigenvalues.

\subsection{Machine Learning Expected Returns}

What we have in mind is a situation where the robo-investors rely on a modeling department within the brokerage house to provide them with estimates of conditional means and conditional covariances for the entire universe of stocks/ETFs and supplying the Alter Ego investor associated with each individual investor with the estimates $\mu_t^i$ and $\Sigma_t^i$ for the stocks in the set $S_{it}.$
The modelers estimate a wide class of models and use out-of-sample performance metrics to determine the most appropriate panel of conditional means and conditional covariances to supply to the robo-investors. Our goal here is to provide a simple approximation to the comprehensive conditional modeling process that such a brokerage research group would undertake. In terms of expected returns models our analysis shares some of the methods also considered by 
\cite{gu2018empirical}.

For the purpose of our analysis,  let $r_{i,t-} = (r_{i, t}, \cdots, r_{i,t-k+1})'$ be the $k \times 1$ vector of own-lagged stock returns for stock $i$. We have $N=1076$ stocks/ETFs to consider and $T=110$ monthly periods. We use 70\% of the data for training, 20\% of the data as a validation sample (for hyperparameter tuning), and 10\% of the sample for testing out-of-sample performance. To maximize the use of our unique data set, we start building our models using returns data from January 1993 to December 2002 - namely a 10-year sample prior to the start of our individual investor data.

We augment the panel of monthly stock/ETF returns with the five Fama-French monthly factors  (Mkt, SMB, HML, RMW, CMA) as well as their momentum factor (see Ken French website for definitions), and \cite{welch2007comprehensive} predictors: div.\ price ratio, div.\ yield, earnings price ratio, div.\ payout ratio, stock variance, BM DJ stocks, net equity expansion, TBill, long-term yield, term spread, default yield spread, inflation (see their paper for definitions).  We understand that a true data engineering group would likely create a much larger and more robust set of data sources. Our goal is not to replicate the true data-source generating process, but to provide a simple approximation to the set of all useful signals for prediction. Let $x_t$ represent an $M \times 1$ vector of these predictors.

In each model class we estimate individual models for each stock/ETF separately, rather than pooling across stocks/ETFs, in order to allow as much heterogeneity as possible in model parameter estimates. The common modeling objective is to estimate $E_t[r_{i,t+1}]$, where $E_t[r_{i,t+1}] = f_i(z_{i,t})$. The modelers therefore employ different approaches to estimate $f_i()$, and also work to curate the best possible set of covariates $z_{i,t}$.

We separate our conditional mean models into (1) linear and (2) nonlinear model sets. Within the linear models, we consider OLS and elastic-net models. For nonlinear models we consider random forests of regression trees and shallow feed-forward neural networks. Hence, we consider two popular nonparametric and parametric machine learning models designed to introduce nonlinear interactions between covariates: random forests of regression trees (nonparametric) and artificial neural networks (parametric). Finally we consider a simple model ensemble across all models.

\paragraph{Linear Models}

The linear models we estimate for each stock $i$ across time periods $t=k,\dots,T-1$ are of the form:
\begin{equation}
\label{eq:linreg}
r_{i,t+1} = \beta_{i,0} + \beta_{i,r} r_{i,t} + \beta_{i,x} x_t + \epsilon_{i,t+1}
\end{equation}
In addition to estimating this model with OLS, we fit sets of linear models per stock $i$ using Elastic Net involving two tuning parameters $(\alpha, \lambda)$ that we optimize over the validation sample.
\begin{equation}
\label{eq:EL}
{\cal L}_i(\theta) = \frac{1}{T-k} \sum_{t=1}^T \epsilon_{i,t+1}^2 + \alpha \lambda \sum_{m \in \# \beta}|\beta_m| + \frac{1}{2}(1-\alpha) \lambda \sum_{m \in \# \beta}\beta_m^2
\end{equation}where $\beta$ = $(\beta_{i,0} \beta_{i,r} \beta_{i,x})^\prime$

\paragraph{Nonlinear Models}

We consider two popular nonparametric and parametric machine learning models designed to introduce nonlinear interactions between covariates: random forests of regression trees (nonparametric) and artificial neural networks (parametric). We employ the algorithm of (\cite{breiman2001random}) to estimate random forest models and we use stochastic gradient descent to minimize an $\ell_2$ objective function with regularization terms in order to train the neural networks. In both cases our estimation techniques are standard. Again we estimate the model on the training data and optimize all respective tuning parameters on the validation set.

\paragraph{Random Forest}

A random forest is a combination of individual regression trees. It is a bootstrapping method that seeks to avoid both overfitting and decrease correlation among trees by using random subsets of predictors at each branch of a given tree. Each tree can be classified as having $K$ terminal nodes (called ``leaves'') with a depth of $L$. The prediction of a given tree then can be stated as:

\begin{equation}
h(z_{i,t}; \beta, K, L) = \sum_{k=1}^K \beta_k 1\{z_{i,t} \in P_k(L)\}
\end{equation}
where $P_k(L)$ is the $k$-th partition that has at most $L$ different branches that it considers. A set of branches for a given partition can be represented as a product of indicators for sequential branches. For a given partition, then $\hat \beta_k$ is the average of the returns for all members of that given partition.
A standard greedy search algorithm is used to maximize the information gained at each split. The recursive binary splitting algorithm continues until a set of stopping criterion are met, which typically rely on the maximal additional information gained from a split being less than a threshold, or a max number of leaves and/or depth of a tree being reached. 

For the random forest models, the key tuning parameters are the number of bootstraped trees, the depth of each tree, and the random subset of predictors that are considered at each potential split within a tree. The random forest prediction is then the bootstrapped average at any prediction point across trees.

\paragraph{Neural Network}

Our neural network architecture is two hidden layers with 10 neurons per layer, sigmoid transfer functions in the input and hidden layers, and a linear transfer function in the output layer. We use stochastic gradient descent to minimize an $\ell^2$ objective function with regularization terms in order to train the neural networks. In both cases our estimation techniques are standard. Again we estimate the model on the training data and optimize all respective tuning parameters on the validation set.

Finally we consider a model ensemble of the above linear and nonlinear estimators, restricting ourselves to an equal-weighting scheme across predicted expected returns as to limit introducing additional estimation uncertainty.

\begin{figure}[htbp]
		\begin{center}
 			\includegraphics[scale=0.5]{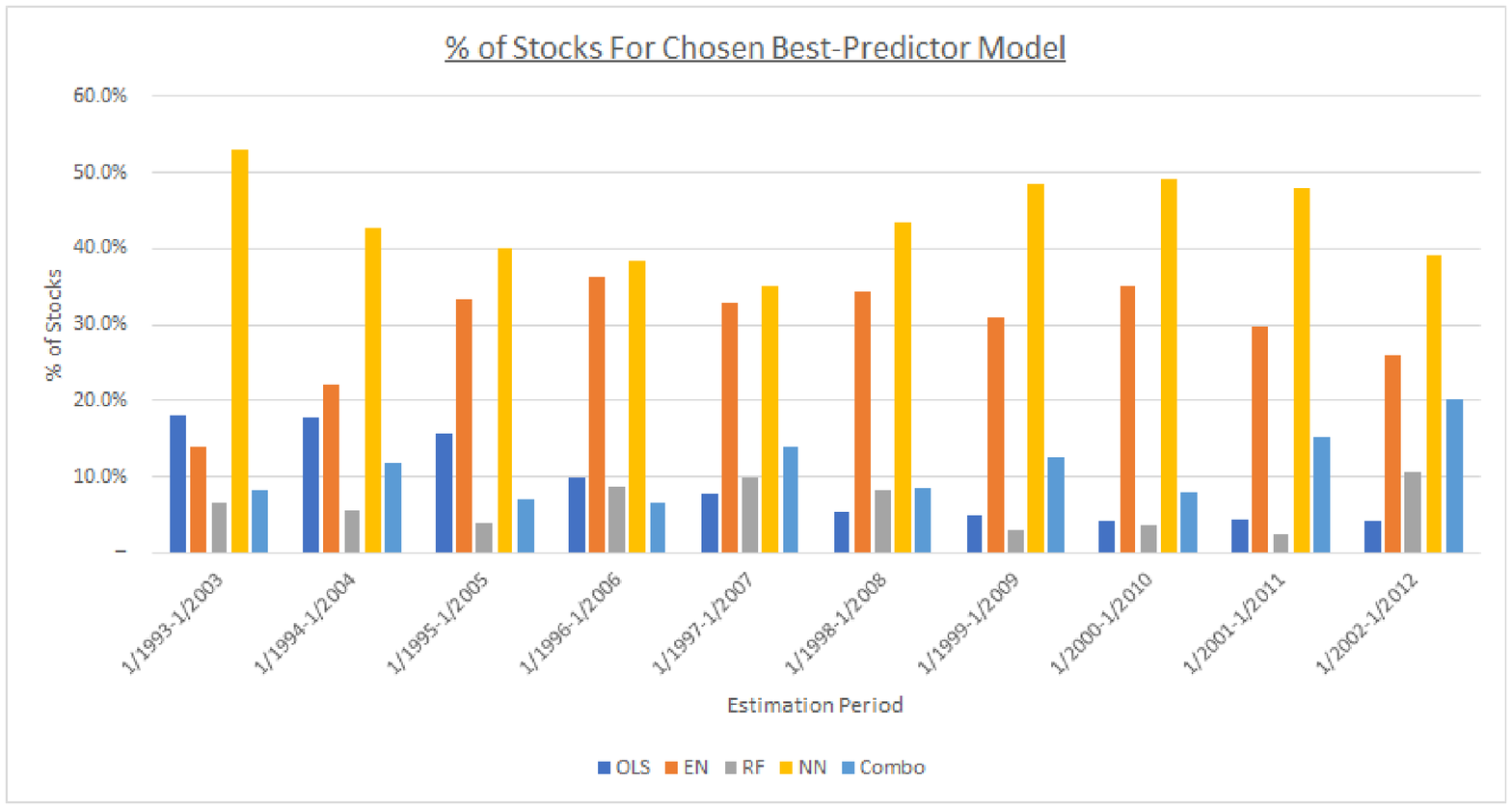}	
		\end{center}
		\caption{ \footnotesize Bar charts for 10-year rolling samples are displayed, where we only display yearly snapshots. The first covers the sample Jan 1993 - Jan 2003 and the last Jan 2002 - Jan 2012.  For each of the 10 rolling samples the relative performance of the competing models (only looking at equities) is displayed.  The bars add up to 100\% for each of the 10 rolling samples. The out-of-sample (OOS) performance is measured in terms of MSE and the height of each bar represents the percentage a particular model has the lowest MSE in predicting the cross-section of returns for all the stocks in the sample. The models are OLS, Elastic Net (EN), Random Forest (RF), Neural Net (NN) and Ensemble (Comb). We use 70\% of the data for training, 20\% of the data as a validation sample (for hyperparameter tuning), and 10\% of the sample for testing OOS performance. The bar charts pertain to the OOS performance.}
		\label{fig:barchartsmodels}	
	\end{figure}

Figure 	\ref{fig:barchartsmodels} displays a set of 10 bar plot clusters.  Each displays end-of-year (last quarter) snapshots of forecasting performance. The 10 rolling samples displayed, each pertaining to a 10-year sample of return data to estimate, validate and forecast returns. For each of the 10 rolling samples the relative performance of the competing models (only looking at equities) is displayed.  The out-of-sample performance is measured in terms of MSE and the height of each bar represents the percentage a particular model has the lowest MSE in predicting the cross-section of returns for all the stocks in the sample. For each cluster the height of the bars add up to 100\% and each represents the fraction a particular class of models provides the best return prediction for the 683 in the cross-section. We note that neural network models represent the most successful class of models, typically being the best for between 40 and 50 percent of the assets in the cross-section. Often a close second is the class of Elastic Net models. All other methods are less successful, although there is quite some variation across time.

The results displayed in Figure \ref{fig:barchartsmodels} may leave the impression that neural network models are dominant. Let us turn our attention to Table \ref{tab:mse} which sheds perhaps a different light on this result. 
Table \ref{tab:mse} reports the average MSE and MAE of out of sample forecasts across all assets and rolling sample schemes. It shows that the elastic-net and neural net models deliver the lowest out-of-sample MSE when aggregating performance across stocks/ETFs. However, the differences between EN and NN are very small, indicating that while NN perhaps provides the best predictions, EN is typically a close second and arguably much easier to implement. Moreover, the EN is a linear model, whereas the NN is nonlinear. The presence of nonlinearities does not seem to substantially pay off.

All the models/estimators have dimensions on where they could be refined, but ultimately the modeling group delivers a set of conditional mean estimates by stock/ETFsto the robo-investors. Each of these chosen conditional mean estimates come from the model with the lowest out-of-sample MSE. A common model need not be chosen across stocks/ETFs, and indeed we can see that even in a few cases, OLS with all covariates included is the model with the best out-of-sample performance. The final panel of $(\hat E_t[r_{i,t+1}])_{i,t}$ is used in the robo-investors' optimal portfolio problems.

\section{Empirical Results \label{sec:empres}}

The empirical results focus on answering a number of questions: (a) are more sophisticated models better, (b) who gains from robo-advice, (c) how does robo-investing perform during a major financial crisis, (d) how do AI Alter Egos compare to passive investment schemes and (e) are spreads due to behavioral biases? A subsection is devoted to answering each of these questions. In the main body of the paper we report a summary set of results pertaining to quarterly rebalancing. In the Online Appendix (\cite{DDGR_SSRN}), the monthly and detailed quarterly results are reported.

\subsection{Are more sophisticated models better?}

In Table \ref{tab:overallspreads} we report for all investors in our sample the median, first (Q1) and third quartiles (Q3) of the cross-sectional distribution of return spreads $r_{i,t}^{s}$ = $r_{i,t}^{ae}$ - $r_{i,t},$ considering only equity holdings (left panel) or the entire universe of  683 stocks and 393 ETFs (right panel). The AI Alter Ego schemes are: (a) MV with Rolling Mean/Rolling Variance, (b) Machine Learning (ML) Mean/Rolling Variance - using the methods displayed in Table \ref{tab:mse}, (c) ML Mean/Nonlinear Smoothed Variance, and finally the equally weighted (EW) portfolio scheme. Neither rolling sample mean nor equally weighted portfolios have positive median spreads. Hence, the median shadow robo-investor performs worse than the humans. The highest median spread is obtained from the ML Mean/Rolling Variance, namely 2.93\% per year (equities only) and 3.37\% for the universe of stocks and ETFs. Using the nonlinear smoothing approach to covariance estimation slightly reduces the median return by 15 basis points or even 41 basis points when ETFs are included. While there is a large cross-sectional heterogeneity, judging by the inter-quartile range, we also observe a right shift in the entire distribution. The first quartiles for MV Rolling Mean and EW are 3 percent lower, whereas Q3 is 5 percent lower compared to either type of MV ML. 
All the results reported so far pertain to quarterly portfolio rebalancing. In the Online Appendix, we provide detailed evidence showing that the findings extend to monthly rebalancing. The annual rebalancing yield qualitatively the same findings as well.

Overall, the results clearly show that the ML expected return scheme is superior to any of the two relatively naive and simple robo-investing schemes. Hence, the answer is clearly that more sophisticated models are better. In the remainder of this section we will therefore focus exclusively on the MV ML/Rolling Variance robo-investor AI Alter Egos.

\subsection{Who gains from robo-advise?}

Continuing with quarterly rebalancing and MV ML/Rolling Variance robo-investors, in Table \ref{tab:sunstatstratified} we report the median, Q1, Q3 as well as confidence interval for the median of the cross-sectional distributions of the spreads between AI Alter Ego and individual investor returns, considering the entire universe of  683 stocks and 393 ETFs.\footnote{\label{footnote:confidencemedian} To construct confidence intervals for aggregate summary statistics we do the following. We first randomly sample individuals according to an individual bootstrap method whereby each investor is assumed independent of each other investor, and sample the entire time-series path of each investor to maintain the dependence structure. For each bootstrap repetition we compute the relevant statistic per individual and aggregate the per-individual statistics over all of the sampled investors. Let $\{\tilde \theta_r \}_{r=1}^R$ be the constructed statistic over $R$ bootstrap repetition, and let $\hat \theta$ be the point estimate of interest. Let $\tilde \theta_{(\alpha/2)}$ and $\tilde \theta_{(1-\alpha/2)}$ represent the $\alpha/2$ and $1-\alpha/2$ percentiles of the bootstrap statistic. We then construct pivotal $1-\alpha$ confidence intervals according to $[2 \hat \theta - \tilde \theta_{(1-\alpha/2)}, 2 \hat \theta - \tilde \theta_{(\alpha/2)}]$.}   Summary statistics are computed for separate samples with low/high education, low/high risk aversion and low/high income classification for investors.

Let us start with high and low risk aversion, which is the panel in the middle of the Table. High risk averse median individual investors stand to gain 5.14 percent from robo-investor shadow Alter Egos. Their low risk aversion counterparts only gain 3.29. Both clearly benefit, since the confidence intervals for either type of investor indicates that the median spreads are significantly different from zero. In addition, the 95\% confidence interval for the difference in medians is $[0.5546, 3.1111],$ and therefore excludes zero. Hence, the median high risk averse investor gains statistically significantly more from robo-investing than the median low risk averse investor does. A similar pattern emerges for high/low income, with the median low income investor gaining roughly two-thirds more (4.13 percent versus 2.76) than the high income median investor. Low and high education differences are not as pronounced, with a wedge of 63 basis points. The inference indicates, however, that the high/low median spread for income and education are not statistically significant. Needless to say that a spread between median 2.76 (high income) and 4.13 (low income) percent return per year is economically quite substantial.


\subsection{How does robo-advising perform during major financial crisis?}

In Table \ref{tab:crisis1} we report the Alter Ego return spreads for stocks/ETFs as they relate to the financial crisis and Great Recession financial. The subsamples are benchmarked using the NBER chronology identifying the crisis period as 12/2007 - 6/2009.\footnote{We also examined the more specifically targeted Belgian crisis dates related to the severe difficulties of the country's financial sector. The results are broadly speaking similar and not reported here.} The focus is again on the MV ML/Rolling Variance  AI Alter Ego scheme.  For each of the subsamples we compute the median, Q1 and Q3 realized returns along with the same statistics for the AI Alter Ego returns. Note that, since the median of a spread is not the difference in median returns, we are not inferring something directly related to the spreads reported in prior tables. We focus on the returns instead in order to highlight a very important finding. Prior to the crisis we note that the median investor had an annual return of 9.26\%, almost double the return of the median AI Alter Ego (4.17\%). We also note though that the inter-quartile spread for investors is twice as large as the same statistic for robo-investors using ML. For individual investors the Q1-Q3 spans from -4.70 to 20.98 percent, whereas the AI Alter Egos feature a better Q1 of minus two percent, and a lower Q3 of almost eleven percent.

During the crisis things take a dramatic turn. The median robo-investor has zero return - meaning the median AI Alter Ego holds cash. In contrast, for individual investors the median is a 29 percent loss and even the Q3 investor still has a -3.59\% negative annual return. Compare that with 23.81 percent return for Q3 of the AI Alter Egos.

After the crisis, things reverse to the pattern observed prior to the crisis - namely the median investor does better than the median AI Alter Ego, with again a much wider inter-quartile range for individual investors, even more than double the dispersion among robo-investors.

\begin{figure}[htbp]
		\begin{center}
 	\includegraphics[scale=0.4]{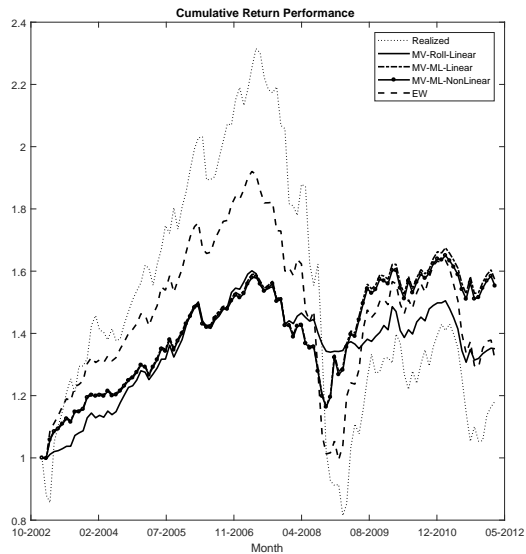}
		\end{center}
		\caption{ \footnotesize The lines correspond to (a) realized return of median investor, (b) median AI Alter Ego returns using MV Rolling Mean/Rolling Variance scheme (c)  median AI Alter Ego returns using MV ML/Rolling Variance scheme (d)  median AI Alter Ego returns using MV ML/Nonlinear Variance and finally (e) the median EW robo-investor. All start out with one unit of investment at the beginning of the sample and median returns are compounded.}
		\label{fig:crisis}	
	\end{figure}

A more striking picture emerges when we turn our attention to Figure \ref{fig:crisis}. The five lines correspond to (a) realized return of median investor, (b) median AI Alter Ego returns using MV Rolling Mean/Rolling Variance scheme (c)  median AI Alter Ego returns using MV ML/Rolling Variance scheme (d)  median AI Alter Ego returns using MV ML/Nonlinear Variance and finally (e) the median EW robo-investor. One word of caution: these medians do not represent the same investor or AI Alter Ego through time, so this is not the performance of a specific individual or robot. Each line starts out with one unit of investment at the beginning of the sample and the median returns are compounded subsequently. Prior to the crisis, the median investor reaches roughly 2.3. This means that the initial capital is doubled over a five year span from 2002 until 2007. By the time the devastation of the crisis took its toll, the median investor is under water by 20 percent and finally ends up with a meager 20 percent return over a 10-year period. It is remarkable that even the EW robo-investor, whom we know from prior analysis is neither sophisticated nor particularly successful, achieves a higher return at the end of the sample. The best overall performance is obtained from the MV ML/Rolling Variance median robo-investor (again not shadowing always the same investor across time) with a 60 percent overall return. This median robo-investor has a relatively slow start and under-performs prior to the crisis, but features small losses during the tumultuous market conditions. Note also that the MV ML/Nonlinear Variance AI Alter Ego is almost identical to the ML/Rolling Variance scheme. Finally, the  MV Rolling Mean/Rolling Variance scheme tracks the ML performance very closely until the financial crisis.

To shed further light on this we turn our attention to Table \ref{tab:ranksEN} displaying the ranking of the regressors based on their $\ell_2$ contribution across stocks for the Elastic Net regressions defined in equations (\ref{eq:linreg}) - (\ref{eq:EL}). We focus on the EN regressions as they provide a fairly simple regression-based interpretation. In addition, it is often the best or nearly the best prediction model. The ranks are computed for 10-year rolling samples starting with 93-03 and ending with 02-12. Of particular interest is the crisis period spanning across the 96-06 through 99-09 samples. The top ranked predictor in all but the last of rolling samples is $dfy$ namely the default yield spread. Another top-ranked series during the crisis is $lty$ or the long term yield. Looking across all samples we also see  $svar$ stock variance, $ntis$ net equity expansion and $infl$ inflation. Interestingly, the usual Fama-French regressors rarely appear among the top-ranked regressors. This should not perhaps come as a surprise, since the Fama-French factors are meant to price the cross-section of returns.

\begin{figure}[htbp]
		\begin{center}
 \includegraphics[scale=.4]{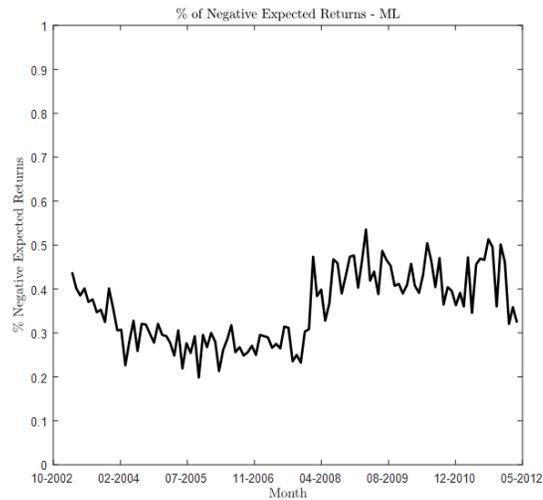}
		\end{center}
		\caption{ \footnotesize Time series plot of the fraction among the cross-section of stocks with negative expected returns, according to the best machine learning model - see Figure \ref{fig:barchartsmodels} for details.}
		\label{fig:fractions}	
	\end{figure}

Finally, in Figure \ref{fig:fractions} we provide a time series plot of the fraction with negative expected returns among the cross-section of stocks, according to the best machine learning model. Early in the sample we see that typically between 20 and 30\% of the stocks featured negative expected returns. The fraction shoots up above 50\% in 2008 and goes as high as 60\%. As a result, the majority of stocks featured negative expected returns, which explains why the AI Alter Egos have a propensity to move out of the market.

\subsection{AI Alter Egos versus Passive Investments}

How do AI Alter Egos measure up against passive investment strategies, in particular buying and holding a market-wide ETF? To address this question we turn our attention to Table \ref{tab:aispreadvsetf}. We report summary statistics for spreads with respect to two ETFs. One tracks the S\&P 500 index and the other is the iShares MSCI Belgium ETF. Neither is ideal, but we did not find an index available throughout the entire sample period that mimics the basket of stocks held by the investors in the brokerage data set.\footnote{Investors hold 26\% of US stocks and 14\% of Belgian stocks.} Unfortunately, the results reported in Table \ref{tab:aispreadvsetf} depend on which ETF is selected.  In the right panel displays the results for stocks+ETFs returns minus the benchmark ETF spreads, either S\&P 500 or Belgian and in the left panel AI Alter Ego MV/ML/Nonlinear against the same benchmarks. Each panel contains the median, first and third quartile of the spreads. The full sample results appear in the top part of Table \ref{tab:aispreadvsetf}. Subsamples stratified according to NBER crisis dates appear in the lower part. The median investor has a spread of -8.50\% against the S\&P 500, meaning the median investor vastly under-performs the benchmark. For the Belgian ETF the results are not as dramatic, since the median investor does better with a positive spread of 1.37\%. There is wide cross-sectional variation, although the third quartile for the US market index is only 2\% (while 12\% for the Belgian index). The AI Alter Ego spreads are better in both cases, although the US benchmark still yields a negative spread of -6.18\%. Against the Belgian ETF, the AI Alter Ego has a positive median spread of almost 4 percent.

When we look at the pre-crisis sample we note that the median investor and AI Alter Ego have returns below the two benchmarks, more so for the Belgian ETF than its US counterpart. It is also worthwhile noting that the median AI Alter Ego performs worse. The crisis period is a totally different story. The AI Alter Egos median investors vastly outperform the benchmark by respectively 18.32\% (SPDR) and 48.31\%. Moreover, the median investor does better than the Belgian ETF by a substantial margin of 21.24\% but is 8.78\% below the S\&P 500 ETF. In both cases we see significant improvements from the Alter Ego schemes. Post-crisis things return back to the pre-crisis situation.

\subsection{Are the spreads due to market or behavioral factors?}

Are the sharp findings regarding the crisis related to well documented behavioral biases? In the Appendix (see \cite{DDGR_SSRN}) we report that the investors in our sample feature the behavioral biases studied in the literature. Regarding the crisis results, we would like to focus on two key ones: (1) the disposition effect (DE) - selling winners too soon, holding on to losers too long - as in \cite{odean1998investors} for each investor and (2) trading frequency.

In Table \ref{tab:dispeff} we document the results of cross-sectional median regressions, where the AI Alter Ego spreads from the MV ML/Rolling Variance robo-investors are a function of DE (left column) as well as DE combined with trading frequencies. We report the AI Alter Ego return spreads as well as the spreads vis-\`a-vis the S\&P 500 ETF. The DE has a positive impact on the spreads, albeit not always statistically significant when combined with trading frequency indicator regressors. When we add dummies for the 2nd through 4th quartile of trading frequency we note that the DE spreads are affected in a statistically significant way, monotonically deteriorating for spreads and increasing for spreads vis-\`a-vis the S\&P 500 ETF. This means that investors who trade a lot tend to have larger AI spreads against the ETF benchmark (in unreported results it is also the case for the Belgian ETF). Conversely, frequent traders tend to have less benefits from AI Alter Egos, and DE is insignificant when combined with trading frequency. Overall, the results indicate that behavioral biases explain to a certain degree the cross-section of AI Alter Ego spreads. Particularly, spreads against a passive investment strategy increase with trading frequency and disposition effect. Additional results involving controls for investor characteristics appear in the Online Appendix. Besides adding controls, we also consider quantile regressions for 5\%, 10\% 20\%, 80\%, 90\% and 95\%. Overall the findings remain, particularly for the right tail of the distribution. The DE is significant for the median and extreme right tail when looking at AI Alter Ego spreads with respect to the benchmark ETF.

\section{Conclusions \label{sec:concl}}

Artificial intelligence enhancements are increasingly shaping our daily lives. Financial decision-making is no exception to this. We introduce the notion of AI Alter Egos, machine-driven decision makers which shadow a particular individual, and apply it in the area of robo-investing using a brokerage accounts data set rich in both cross-sectional and time series features. 

The purpose of our analysis is to assess the highly touted benefits of robo-advising. Through the AI Alter Ego scheme we address a number of questions: (a) are more sophisticated models better, (b) who gains from robo-advise, (c) how does robo-investing perform during a major financial crisis, (d) how do AI Alter Egos compare to passive investment schemes and (e) are spreads due to behavioral biases?  Overall, we find that investors displaying certain characteristics - in particular high risk averse and low income - stand to gain significantly. In particular: high risk-aversion, low income investors. Moreover, machine learning methods provide important portfolio return improvements. AI Alter Ego spreads are related to behavioral biases - in particular the disposition effect and trading frequency. During the financial crisis, robo-investors have a greater propensity to cash out of the market, which contributes to their overall return superiority. Finally, compared to passive ETF investment, we find that the evidence is mixed, although during the financial crisis AI Alter Egos were vastly better than the passive strategy.


\bibliography{bib_DDGR}


\bigskip


\begin{table}[h]
	\begin{center}
		\caption{Out-of-Sample MSE Across Stocks\label{tab:mse}}
		\begin{tabular}{lrrrrr}
			& \multicolumn{5}{c}{Cross-Sectional MSE} \bigstrut[b]\\
			\cline{2-6}          & \multicolumn{1}{c}{OLS} & \multicolumn{1}{c}{EN} & \multicolumn{1}{c}{RF} & \multicolumn{1}{c}{NN} & \multicolumn{1}{c}{Comb} \bigstrut\\
			\cline{2-6}    Mean  & 0.0207  & 0.0104  & 0.0114  & 0.0100  & 0.0101  \bigstrut[t]\\
			Median & 0.0147  & 0.0072  & 0.0081  & 0.0066  & 0.0070  \\
		\end{tabular}%
	\end{center}
	\vskip+5pt \footnotesize{\emph{Notes:}
		Cross-sectional average and median MSE's on the out-of-sample testing data for: OLS, Elastic-Net (EN), Random Forest (RF), Neural Network (NN), and ensemble (Comb).
	}
\end{table}%
\begin{table}[htbp]
	\caption{AI Alter Ego Return Spreads - All Investors 	\label{tab:overallspreads}}
	\begin{center}
		{\small
			\begin{tabular}{lccccccc}
				&  \multicolumn{3}{c}{Equities only}  &     &   \multicolumn{3}{c}{Equities + ETF} \bigstrut[b]\\
					& \multicolumn{1}{c}{Median} & \multicolumn{1}{c}{Q1}  & \multicolumn{1}{c}{Q3} & 	& \multicolumn{1}{c}{Median} & \multicolumn{1}{c}{Q1}  & \multicolumn{1}{c}{Q3} \bigstrut[b]\\
				\cline{2-8}       
	&  &  &  &  &  & &  \\
		 & \multicolumn{7}{c}{Mean Variance} \\
		Rolling Mean/Rolling Variance & -0.08 &	-13.60 	& 12.84 & & 0.58 &	-12.72 &	13.16 \\		
				&  &  &  &  &  & &  \\		
ML Mean/Rolling Variance			 & 2.93 &	-10.66 	& 17.41   &  &   3.37 &	 	-10.19 &	17.20    \\
	&  &  &  &  &  & &  \\
ML Mean/Nonlinear Smoothed Variance 	&    2.78 &  	-10.72 &	17.13   &   &   2.96 &	-10.49 &	17.09 
  \\
	&  &  &  &  &  & &  \\
& \multicolumn{7}{c}{Equally Weighted} \\
&-0.67 &	-13.88 &	11.96 	 & &-0.54  &	-13.68 	& 11.98  \\
			\end{tabular}%
		}
	\end{center}
	\vskip+5pt \scriptsize{\emph{Notes:} Entries are median, first (Q1) and third (Q3) quartiles of the cross-sectional distributions of the  spreads between AI Alter Ego and individual investor returns, $r_{i,t}^{s}$ = $r_{i,t}^{ae}$ - $r_{i,t},$ for three Mean Variance (MV) types of robo-investors and one equally weighted (EW) considering only equity holdings (left panel) or the entire universe of  683 stocks and 393 ETFs (right panel). The AI Alter Ego schemes are: (a) MV with Rolling Mean/Rolling Variance, (b) Machine Learning (ML) Mean/Rolling Variance - using the methods displayed in Table \ref{tab:mse}, (c) ML Mean/Nonlinear Smoothed Variance, and finally the equally weighted portfolio scheme. The spreads are in percentage per year.}
\end{table}%

	\begin{table}[htbp]
	{\small
		\centering
		\caption{Ranked Variables Based on Relative $\ell_2$ Contribution Across Stocks \label{tab:ranksEN}}
		\begin{tabular}{lllllllllll}
			\toprule
			&       &       &       &       &       &       &       &       &       &  \\
			Rank & 93-03 & 94-04 & 95-05 & 96-06 & 97-07 & 98-08 & 99-09 & 00-10 & 01-11 & 02-12 \\
			\midrule
			1     & SMB & svar & ntis & dfy & dfy & dfy & lty & Mkt-RF & ntis & svar \\
			2     & dfy & lty & lty & lty & svar & ntis & dfy & lty & svar & dfy \\
			3     & lty & infl & infl & svar & SPvwx & lty & svar & svar & Mkt-RF & lty \\
			4     & RMW & dfy & tbl & infl & infl & svar & tbl & dfy & infl & infl \\
			5     & infl & Mkt-RF & svar & HML & Mkt-RF & tbl & Mkt-RF & SPvw & SPvw & ntis \\
			6     & ntis & SMB & HML & tbl & dy & Mkt-RF & infl & ntis & CMA & RMW \\
			7     & HML & RMW & RMW & Mkt-RF & lty & infl & ntis & tbl & bm & tbl \\
			8     & Mkt-RF & ntis & dy & bm & RMW & dy & SPvw & CMA & lty & SPvwx \\
			9     & svar & HML & bm & SMB & ntis & SPvwx & RF & bm & ltr & SPvw \\
			10    & SPvw & bm & dfy & SPvw & dp & dp & SPvwx & RMW & SMB & Mkt-RF \\
			11    & Mom & Mom & Mkt-RF & RF & SMB & SPvw & HML & SMB & SPvwx & CMA \\
			12    & tbl & ltr & ep & ntis & tbl & bm & CMA & HML & dfy & HML \\
			13    & ltr & dy & RF & dy & bm & SMB & RMW & SPvwx & HML & SMB \\
			14    & de & SPvw & SMB & CMA & HML & RMW & bm & dy & dy & ltr \\
			15    & bm & CMA & dp & RMW & RF & HML & SMB & Mom & Mom & dp \\
			16    & SPvwx & de & CMA & SPvwx & SPvw & CMA & ltr & dp & dp & RF \\
			17    & RF & SPvwx & SPvw & Mom & Mom & RF & dy & ltr & tbl & Mom \\
			18    & CMA & dp & Mom & ltr & de & de & dp & infl & RMW & bm \\
			19    & dy & tbl & ltr & ep & ep & ep & Mom & ep & ep & dy \\
			20    & ep & RF & de & dp & CMA & ltr & de & de & RF & ep \\
			21    & dp & ep & SPvwx & de & ltr & Mom & ep & RF & de & de \\
			&       &       &       &       &       &       &       &       &       &  \\
		\end{tabular}%
	\vskip+5pt \footnotesize{\emph{Notes:} Elastic Net regressions defined in equations (\ref{eq:linreg})-(\ref{eq:EL}) involve the following set of regressors: $dp$ Dividend/Price, $dy$ Dividend Yield,
$ep$ Earnings/Price, $de$ Dividend Payout, $svar$ Stock Variance, $bm$ Book-to-Market, $ntis$ Net Equity Expansion, $tbl$ T-Bill Rate, $lty$ Long Term Yield, $ltr$ Long Term Return,  $dfy$ Default Yield Spread,
$infl$ Inflation, $SPvw$ S\&P 500, $SPvwx$ S\&P 500 (excl.\ dividends), the following Fama French factors $Mkt-RF$ Market, $SMB,$ $HML,$ $RMW,$ $CMA,$ $RF,$ and $Mom$ Momentum (see Ken French website for definitions), and \cite{welch2007comprehensive} for definitions.}
	}
	\end{table}%


\begin{table}[htbp]
	\caption{AI Alter Egos Return Spreads - Education, Risk Aversion and Income \label{tab:sunstatstratified}}
	\begin{center}
		{\small
				\begin{tabular}{lrrrccc}
					& \multicolumn{3}{c}{Return Spreads}   &       & \multicolumn{2}{c}{CI Median} \bigstrut[b]\\
					\cline{2-7}       
					& \multicolumn{1}{c}{Median} & \multicolumn{1}{c}{Q1}  & \multicolumn{1}{c}{Q3} & & \multicolumn{1}{c}{C(2.5\%)} & \multicolumn{1}{c}{C(97.5\%)} \bigstrut\\
					\cline{2-7}   
					&  &  &  &  &  &      \\
					\textit{Education} &       &   &    &       &     &     \\
					Low   &  2.83 & 	-10.47 	& 16.79 
					& &	 1.18 &	 4.22   \\
					High  & 3.46 & 	-10.15 	& 17.26  & & 3.01 &	3.85	 \\
					&       &       &       &   &    &              \\
					 \multicolumn{7}{l}{Confidence interval difference in medians: [-0.5989,	2.1723] } \\
					 	&       &       &       &   &    &              \\
					\textit{Risk Aversion} &    &   &       &       &   &    \\
					Low &  3.29 & 	-10.86 &	17.43 
					& &  2.39 & 	4.38    \\
					High & 5.14 &  	-9.12 &	17.28   & &  3.63 & 	 6.36 \\ 
					&       &       &       &   &    &              \\
				\multicolumn{7}{l}{Confidence interval difference in medians: [0.5546,	3.1111] } \\
				&       &       &       &   &    &              \\
					\textit{Income} &       &    &   &       &      &     \\
					Low & 4.13 & 	-10.00 & 	17.57  & & 3.01 & 5.20 
					\\
					High & 2.76 & 	-11.70 &	15.01 
					& &  0.82 & 4.91 \\
					&       &       &       &   &    &              \\
				\multicolumn{7}{l}{Confidence interval difference in medians: [-3.1016, 0.6014] } \\
				&       &       &       &   &    &              \\	
				\end{tabular}%
		}
	\end{center}
	\vskip+5pt \footnotesize{\emph{Notes:} Entries are median, first and third quartiles, as well as confidence interval for the median of the cross-sectional distributions of the spreads between AI Alter Ego and individual investor returns, considering the entire universe of  683 stocks and 393 ETFs. The AI Alter Ego scheme is Mean Variance (MV) with Machine Learning (ML) Mean/Rolling Variance - using the methods displayed in Table \ref{tab:mse}. Summary statistics are computed for separate samples with low/high education, low/high risk aversion and low/high income classification for investors. 95\% confidence intervals for differences in medians are computed as described in footnote \ref{footnote:confidencemedian}. The spreads are in percentages per year.}
\end{table}%

\begin{table}[htbp]
	\caption{Returns Pre-Crisis, Crisis and Post-Crisis \label{tab:crisis1}}
	\begin{center}
		{\small
			\begin{tabular}{lrrr}
				& \multicolumn{1}{c}{Median} & \multicolumn{1}{c}{Q1} & \multicolumn{1}{c}{Q3} \\
				\cmidrule{2-4}  
				& & & \\
				Pre & & & \\
				Realized  &  9.26 &	-4.70 &	20.98  \\
				MV ML &	4.17 & -2.00 &	10.70  \\ 	
				& & & \\
				During & & & \\
				Realized  & -29.04 &	-45.87 	& -3.59  \\
				MV ML &	0.00 & -13.81 	& 23.81  \\    
				& & & \\
				Post & & & \\
				Realized  &  5.64  & -6.47  & 15.30  \\
				MV ML &		2.05 &	-1.76 &	8.42 \\		
			\end{tabular}	
		}
	\end{center}
	\vskip+5pt \footnotesize{\emph{Notes:} The subsamples are benchmarked based on the NBER Crisis Time Period 12/2007 - 6/2009. The Pre-crisis sample starts in 2002 and ends 11/2007, the post-crisis sample covers 7/2008 until end of sample, 2012.  MV ML refers to the AI Alter Ego scheme is Mean Variance with Machine Learning Mean/Rolling Variance - using the methods displayed in Table \ref{tab:mse}. The spreads are in percentages per year.}
\end{table}%

\begin{table}[htbp]
	
	\caption{AI Alter Ego Return Spreads vis-\`a-vis benchmark ETFs}
	\begin{center}
	\begin{tabular}{lrrcrrrr}
		& \multicolumn{3}{c}{Realized Stocks+ETFs} &       &       \multicolumn{3}{c}{AI Alter Ego}	       \\
		& \multicolumn{3}{c}{minus ETF} &       &      	\multicolumn{3}{c}{minus ETF}        \\
		&       &         &       &       &       &       &  \\
		&      &       &       &       &       &       &  \\
		& \multicolumn{1}{c}{Median} & \multicolumn{1}{c}{25q} & \multicolumn{1}{c}{75q} &   & \multicolumn{1}{c}{Median} & \multicolumn{1}{c}{25q} & \multicolumn{1}{c}{75q}  \\
		\cmidrule{2-8} 	&       &     &       &       &       &    &      \\
		& \multicolumn{7}{c}{Full sample} \\ 	
		&       &       &       &       &       &       &       \\
		S\&P 500 ETF  & -8.50  & -21.05  & 2.00    &    & -6.18  & -13.99  & 1.32  \\
		Belgian ETF	&   1.37 &-9.90 &	12.00   &  &  3.93 & -4.89 & 13.10   \\
		&       &       &       &       &       &       &          \\
		& \multicolumn{7}{c}{Pre-Crisis} \\
		&       &       &       &       &       &       &      \\
		S\&P 500 ETF  & -2.22  & -16.02  & 9.18  &       &  -7.20  & -13.88  & -0.41  \\
		Belgian ETF	& -9.57 & -21.28 &	1.45 & & -14.58 & -21.33 & -7.02  \\
		&       &       &       &       &       &       &       \\
		& \multicolumn{7}{c}{During Crisis} \\
		&       &       &       &       &       &       &        \\
		S\&P 500 ETF  & -8.78  & -27.02  & 12.48  &     & 18.32  & -1.75  & 39.32  \\
		Belgian ETF	& 21.24 & 1.23 	& 40.35 & & 48.31 &	25.60 &	71.02 \\ 
		&       &       &       &       &       &       &         \\
		& \multicolumn{7}{c}{Post-Crisis} \\
		&       &       &       &       &       &       &         \\
		S\&P 500 ETF & -11.37  & -23.16  & -1.78  &   & -14.97  & -19.62  & -8.43  \\
		Belgian ETF	& -1.48 & -12.59 & 8.26  & & -4.82 & -9.85 &	2.36 \\  
	\end{tabular}%
	\end{center}
\vskip+5pt \footnotesize{\emph{Notes:} The AI Alter Ego scheme is the ML Mean/Nonlinear Smoothed Variance  - using the methods displayed in Table \ref{tab:mse}. The spreads are in percentage per year. The subsamples are benchmarked based on the NBER Crisis Time Period 12/2007 - 6/2009. The Pre-crisis sample starts in 2002 and ends 11/2007, the post-crisis sample covers 7/2008 until end of sample, 2012. The benchmark ETFs are the SPDR ETF tracking the S\&P 500 index and the iShares MSCI Belgium ETF.}
	\label{tab:aispreadvsetf}%
\end{table}%

\begin{table}[htbp]
	\caption{Disposition Effect and AI Alter Ego Return Spreads - Median regression	\label{tab:dispeff}}
	\begin{center}
    \begin{tabular}{lccccc}
	&       &   & & &   \\
	& \multicolumn{2}{c}{Spreads} & & \multicolumn{2}{c}{Spreads vis-\`a-vis} \\
		& & & & \multicolumn{2}{c}{ S\&P 500 ETF} \\
			\cmidrule{2-3} 	\cmidrule{5-6}
			&       &   & & &   \\
	&   \multicolumn{1}{c}{Model DE}    &   \multicolumn{1}{c}{Model DE}  & & \multicolumn{1}{c}{Model DE}    &   \multicolumn{1}{c}{Model DE} \\
		&   \multicolumn{1}{c}{Only}     &  \multicolumn{1}{c}{+ Trading Freq.}  & &   \multicolumn{1}{c}{Only}     &  \multicolumn{1}{c}{+ Trading Freq.}   \\
Trading frequency   &  &  & &  & \\
	&       &   & & &  \\
2nd quartile   &       & -1.361* & &       & 0.782*** \\
	&       &    & & &  \\
3rd quartile  &       & -1.944*** & &      & 1.168*** \\
	&       &       & & & \\
4th quartile  &       & -2.850*** & &       & 1.175*** \\
			&       &   & & &   \\
Disposition Effect  & 0.014* & 0.011 &  & 0.004* & 0.007*** \\
   \end{tabular}
	\end{center}
	\vskip+5pt \footnotesize{\emph{Notes:} Cross-sectional Median Regression Alter Ego Spreads with MV ML/Rolling Variance, * p$<$0.10 ** p$<$0.05, *** p$<$0.01. Number of observation $N$ = 19118. Detailed parameter estimates for the controls (gender, education, risk aversion, income, funds invested, ETF use) appear in the Online Appendix.}
\end{table}%

\end{document}